\magnification=1200 \vsize=25truecm \hsize=16truecm \baselineskip=0.6truecm
\parindent=1truecm \nopagenumbers \font\scap=cmcsc10 \hfuzz=0.8truecm
\font\tenmsb=msbm10
\font\sevenmsb=msbm7
\font\fivemsb=msbm5
\newfam\msbfam
\textfont\msbfam=\tenmsb
\scriptfont\msbfam=\sevenmsb
\scriptscriptfont\msbfam=\fivemsb

\null \bigskip  \centerline{\bf AGAIN, LINEARIZABLE MAPPINGS}

\vskip 2truecm
\bigskip 
\centerline{\scap A. Ramani} 
\centerline{\sl CPT, Ecole Polytechnique}
\centerline{\sl CNRS, UPR 14} 
\centerline{\sl 91128 Palaiseau, France}
\bigskip
\centerline{\scap B. Grammaticos}
\centerline{\sl GMPIB (ex LPN), Universit\'e Paris VII} 
\centerline{\sl Tour 24-14, 5$^e$\'etage} 
\centerline{\sl 75251 Paris, France}
\bigskip 
\centerline{\scap K.M. Tamizhmani}
\centerline{\sl Departement of Mathematics}
\centerline{\sl Pondicherry University}
\centerline{\sl Kalapet, Pondicherry, 605014 India}
\bigskip 
\centerline{\scap S. Lafortune$^{\dag}$} 
\centerline{\sl LPTM et GMPIB,  Universit\'e Paris VII} 
\centerline{\sl Tour 24-14, 5$^e$\'etage} 
\centerline{\sl 75251 Paris, France}
\footline{\sl $^{\dag}$ Permanent address: CRM, Universit\'e de
Montr\'eal, Montr\'eal, H3C 3J7 Canada}
\bigskip\bigskip 
Abstract
\medskip\noindent
We examine a family of 3-point mappings that include mappings solvable through linearization. The
different origins of mappings of this type are examined: projective equations and Gambier systems. The
integrable cases are obtained through the application of the singularity confinement criterion and
are explicitly integrated.

\vfill\eject

\footline={\hfill\folio} \pageno=2

\bigskip
\noindent {\scap 1. Introduction}
\medskip 
Integrability is a term far too general. Although the idea is simple and related to the integration
of a differential system, integrability assumes various forms. Here we shall attempt neither a rigorous
definition nor an exhaustive description of all the disguises of integrability. In order to fix the ideas
we shall just present three most common types of integrability, which suffice in order to explain the
properties of the majority of integrable systems [1]. These three types are:

- Reduction to quadrature through the existence of the adequate number of integrals of motion.

- Reduction to linear differential systems through a set of local transformations.

- Integration through IST techniques. This last case is mediated  by the existence of a Lax pair (a
linear system the compatibility of which is the nonlinear equation under integration) which allows the
reduction of the nonlinear equation to a linear integrodifferential one.
The above notions can be extended {\sl mutadis mutandis} to the domain of discrete systems.

This paper will focus on the second type of integrability, usually refered to as linearizability. The
prototype of the linearizable equations is the Riccati. In differential form this equation writes:
$$w'=\alpha w^2+\beta w+\gamma\eqno(1.1)$$
and the transformation $w=P/Q$ (Cole-Hopf) reduces it to the linear system:
$$P'= \beta P+\gamma Q    \eqno(1.2)$$
$$Q'= -\alpha P     $$
Similarly, the discrete Riccati equation, which assumes the form of a homographic mapping:
$$\overline x ={ {\alpha x+\beta}\over {\gamma x+\delta}}\eqno(1.3)$$
where $x$ stands for $x_n$, $\overline x$ for $x_{n+1}$ (and, of course, $\underline x$ for $x_{n-1}$),
can be linearized through a Cole-Hopf transformation. Putting $x=P/Q$ we obtain readily:
$$\overline  P= \alpha P+ \beta Q  \eqno(1.4)$$
$$\overline  Q=\gamma P + \delta Q $$
The extension of the Riccati to higher orders can be and has been obtained [2]. The simplest linearizable
system at $N$ dimensions is the projective Riccati which assumes the form:
$$w'_\mu=a_\mu+\sum_\nu b_{\mu\nu}w_\nu+w_\mu\sum_\nu c_\nu w_\nu\quad{\rm with}\ 
\mu=1,\dots,N\eqno(1.5)$$ In two dimensions the projective Riccati system can be cast into a second
order equation without loss of generality:
$$w''= -3ww'-w^3+q(t)(w'+w^2)\eqno(1.6)$$
The discrete analog of the projective Riccati does exist and is studied in detail in [3]. The
corresponding form is:
$$\overline x_\mu={a_\mu+x_\mu+\sum_\nu b_{\mu\nu}x_\nu\over 1-\sum_\nu c_\nu x_\nu}\eqno(1.7)$$
Again in two dimensions (and only in this case) the discrete projective Riccati can be written as a
second-order mapping for a variable without any simplifying assumptions. One is thus led to a mapping
of the form
$$\alpha\overline x x  \underline x+\beta \overline x x+ \gamma x \underline x +
\delta \overline x \underline x + \epsilon x + \zeta \overline x +\eta \underline x                    
+\theta=0 \eqno(1.8)$$ 
which was first introduced in [4]. The coefficients $\alpha,\beta,\ldots,\theta$ are not totally free.
Although the linearizability constraints have been obtained in [4], the study of mappings of the
form (1.8) was not complete. In the present work we intend to present its exhaustive study from the
point of view of integrability in general.

Another point must be brought to attention here. In the continuous case the study of second order
equations has revealed the relation of the linearizable equations (1.6) to the Gambier equation [5]. The
latter is obtained as a system of two coupled  Riccati in cascade
$$y'=-y^2+qy+c \eqno(1.9)$$
$$w'=aw^2+nyw+\sigma$$
where $n$ is an integer. The linearizable equation (1.6) is obtained from (1.9) for $n=1$ and $a=-1$, $c=0$
and $\sigma=0$.
The discrete analog of the Gambier mapping was introduced in [6] and in full generality in [7]. However
the relation of the linearizable mapping to the discrete Gambier system has not been studied in full
detail. In what follows we shall fill this gap by presenting the reduction of the Gambier mapping  to the
linearizable one.

In the next section we shall analyse (1.8) and all its reduced forms and isolate the integrable ones
through the use of the singularity confinement criterion. The integration of all integrable cases will be
given in detail and in particular the linearization through a projective system. The third section will
be devoted to the study of the reduction of a Gambier-type (coupled Riccati) system to a linearizable
one together with the investigation of the integrable cases and their integration.
\bigskip
\noindent {\scap 2. Linearizable mappings as projective systems}
\medskip
In [4] we have introduced a projective system as a way to linearize a second-order mapping. (The general
theory of discrete projective systems has been recently presented in [3]). In this older work of ours we
have focused on a three-point mapping that can be obtained from a $3\times3$ projective system. As a
matter of fact, this is the only case where the projective equation can be converted to a single, one 
component, mapping without any simplifying assumptions. The main idea was to consider the system:
$$\pmatrix{\overline u\cr\overline v\cr \overline w}=
\pmatrix{p_{11}&p_{12}&p_{13}\cr p_{21}&p_{22}&p_{23}\cr p_{31}&p_{32}&p_{33}}
\pmatrix{u\cr v\cr w}\eqno(2.1)$$
and conversely
$$\pmatrix{\underline u\cr\underline v\cr \underline w}=
\pmatrix{m_{11}&m_{12}&m_{13}\cr m_{21}&m_{22}&m_{23}\cr m_{31}&m_{32}&m_{33}}
\pmatrix{u\cr v\cr w}\eqno(2.2)$$
where the matrix $M$ is obviously related to the matrix $P$ through $\overline M=P^{-1}$ . Introducing
the variable
$x=u/v$ and the auxiliary $y=w/v$ we can rewrite (2.1) and (2.2) as 
$$\overline x={p_{11}x+p_{12}+p_{13}y\over p_{21}x+p_{22}+p_{23}y}\eqno(2.3)$$
$$\underline x={m_{11}x+m_{12}+m_{13}y\over m_{21}x+m_{22}+m_{23}y}\eqno(2.4)$$
(Since the $m_{3i}$, $p_{3i}$ do not appear in (2.3), (2.4) and we can simplify
$M,P$ by taking $m_{33}=p_{33}=1$ and $m_{31}=p_{31}= m_{32}=p_{32}=0 $).
Finally eliminating $y$ between (2.3) and (2.4) we obtain the mapping:
$$\alpha\overline x x  \underline x+\beta \overline x x+ \gamma x \underline x +
\delta \overline x \underline x + \epsilon x + \zeta \overline x +\eta \underline x                    
+\theta=0 \eqno(2.5)$$ 
where the $\alpha, \beta, \ldots \theta $ are related to the $m, p$'s. 

Equation (2.5) will be the starting point of the present study. Our question will be when is an equation
of this form integrable? (Clearly the relation to the projective system works only for a particular
choice of the parameters). In order to investigate the integrability of (2.5) we shall use the
singularity confinement approach that was introduced in [8]. The first question is what are the
singularities of (2.5)? Given the form of (2.5) it is clear that diverging $x$ does not lead to any
difficulty. However, another (subtler) difficulty arises whenever $x_{n+1}$ is defined independently of
$x_{n-1}$. In this case the mapping ``loses one degree of freedom''. Thus the singularity condition is
$${\partial x_{n+1} \over \partial x_{n-1}}=0$$
which leads to :
$$(\alpha  x+\delta)(\epsilon x+\theta)=(\beta x+\zeta)(\gamma x+ \eta)  \eqno(2.6)$$
Equation (2.6) is the condition for the appearence of a singularity. Given the invariance of (2.5) under
homographic transformations it is clear that one can use them in order to simplify (2.6). Several choices
exist but the one we shall make here is to choose the roots of (2.6) so as to be equal to $0$ and
$\infty$, unless of course (2.6) has two equal roots, in which case we bring them both to $0$. Let us
examine first the distinct root case. For the roots of (2.6) to be $0$ and $\infty$ we must have:
$$\alpha \epsilon    =\beta\gamma \eqno(2.7)$$
$$\delta\theta=\zeta\eta$$
The generic mapping of the form (2.5) has $\alpha\theta\neq0$ and we can take $\alpha=\theta=1$ (by
the appropriate scaling of $x$ and a division). We have thus,
$$\overline x x  \underline x+\beta \overline x x+ \gamma x \underline x +\zeta\eta \overline x
\underline x + \beta\gamma x + \zeta \overline x +\eta \underline x +1=0\eqno(2.8)$$

Nongeneric cases do exist as well and we shall examine them in detail. Thus let us first assume that
$\alpha =0$ in which case $\beta\gamma=0$ and we can decide that $\beta=0$ without loss of
generality ($\beta$  and  $\gamma$ are interchanged if one reverses the direction of the evolution).
The mapping then becomes: $(\gamma \underline x + \epsilon )x+(\zeta \overline x+1)(\eta \underline
x+1)=0$ and we must have $\zeta \neq 0$, lest the mapping become a two-point one, in which case a scaling
of $x$ can bring its value to 1. We have thus the general form:
$$ x(\gamma\underline x+\epsilon) + (\overline x+1)(\eta\underline x+1)= 0 \eqno(2.9)$$
Clearly we cannot take both $\eta=0,\gamma=0$, neither $\epsilon=0,\gamma=0$. So, in order to obtain the
reduced forms of (2.9) we assume first $\eta=0$ leading to
$$\overline x+1+x(\gamma\underline x+\epsilon)=0 \eqno(2.10)$$
and then $\eta \neq0$ with two possible combinations either $\gamma=0$: 
$$(\overline x+1)(\eta\underline x+1)+\epsilon x=0 \eqno(2.11)$$
or $\epsilon=0$
$$(\overline x+1)(\eta\underline x+1)+\gamma   x\underline x =0 \eqno(2.12)$$
Going back to mapping (2.5) we can obtain a further reduced form by assuming $\alpha=\theta=0$, in which
case $\beta\gamma =0$ and $\zeta\eta=0$.
As previously, we can assume that $\beta=0$ but this fixes the direction of evolution so the two
choices $\zeta=0$ and $\eta=0$ are distinct. We have thus two reduced mappings:
$$\gamma x \underline x + \delta\overline x \underline x +\epsilon x +\zeta \overline x =0 \eqno(2.13)$$
and
$$\gamma x \underline x + \delta\overline x \underline x +\epsilon x +\eta \underline x =0 \eqno(2.14)$$
A further reduction can be obtained if we assume that $\beta=\gamma=0$, in which case we can freely
choose
$\zeta=0$ and find:
$$\delta\overline x \underline x +\epsilon x +\eta \underline x =0 \eqno(2.15)$$
Mappings (2.8) to (2.15) are all that can be obtained from (2.5) with the assumptions of the distincts
roots for (2.6) up to homographic transformations. 

Next we turn to the case where (2.6) possesses a double
root equal to 0. The constraints in this case read:
$$\delta \theta = \zeta \eta \eqno(2.16)$$
$$\alpha \theta + \delta \epsilon = \beta \eta + \zeta \gamma$$
The generic case corresponds to $\theta = \epsilon = 1$, leading to :
$$\alpha\overline x x  \underline x+\beta \overline x x+ \gamma x \underline x +
\zeta \eta\overline x \underline x + x + \zeta \overline x +\eta \underline x                   
+1=0 \eqno(2.17)$$
with  $\alpha = \beta \eta + \zeta \gamma-\delta $.
Nongeneric cases exist also and we start by considering $\theta=0$. This leads to
$\zeta \eta =0$ and we choose $\zeta=0$. One further constraint must be satisfied in this case: $\delta
\epsilon= \beta \eta$. Assuming $\epsilon=1$, we have $ \delta= \beta \eta$ and obtain:
$$\alpha \overline x x \underline x + \beta \overline x x + \gamma x \underline x + \beta \eta \overline
x \underline x + x +\eta \underline x =0 \eqno (2.18)$$
If we take $ \epsilon=0$ then two choices exist: $\eta=0$ or $ \beta=0$.
The first leads to:
$$\alpha \overline x x \underline x + \beta \overline x x + \gamma x \underline x + \delta \overline
x \underline x =0 \eqno (2.19)$$
while the second gives:
$$\alpha \overline x x \underline x + \gamma x \underline x +\delta\overline
x \underline x +  \eta \underline x =0 \eqno (2.20)$$
We can immediately point out that the mappings (2.19) and (2.20) are trivial. The first reduces to a
affine one under the transformations $x\to 1/x$, while in the second the term $\underline x$ is a
common factor and the mapping is in fact a two-point one.
The mappings (2.17) to (2.20)  complete the reductions of (2.5) in the case of a double root of (2.6).

In order to investigate the integrability of the mappings obtained above we shall apply the singularity
confinement criterion as previously explained. We start with the generic mapping (2.8). 
Here the singularities are by construction 0 and $\infty$. Following the results of [4] we require
confinement in just one step. These leads to the condition  $\beta=\zeta=0$. We obtain thus the mapping:
$$\overline x x \underline x + \gamma x \underline x+ \eta \underline x + 1 =0 \eqno(2.21)$$
or, solving for $\overline x$ :
$$\overline x = - \gamma + {\eta \underline x + 1\over x \underline x} \eqno(2.22)$$
where $\gamma$ and $\eta$ are free. We expect 
(2.22) to be integrable. This is indeed the case: we can show that (2.22) can be obtained from the
projective system (2.1), (2.2) provided we take
$$P=\pmatrix{-(\gamma q+1)&q \overline q&1\cr q&0&0\cr0&0&1}
\eqno(2.23)$$
and $M=\underline P^{-1}$ provided $q$ is some solution of the equation $ \overline q q
\underline q + q\underline q \eta + \underline q\, \underline \gamma + 1= 0 $. Then we can use the
projective system (2.1) and (2.2) in order to construct the general solutions of (2.21). (Let us point
out here that the equation for $q$ is really different from (2.21) due to the presence of $\underline
\gamma$ rather than $\gamma$).

We turn now to the remaining mappings starting with (2.9).
For the application of the singularity criterion we must find the values of $x$ for which the mapping
loses one degree of freedom. For mapping (2.9) this happens (by construction) when $x$ is 0 or $\infty$,
but another source of singularity exists, when $\gamma {\underline x}+\epsilon=0$. The investigation of
the singularity $x=0$ leads to the confinement condition $\epsilon=\eta=1$. Proceeding now to the
examination of the singularity starting from $\gamma {\underline x}+1=0$ we find that there is no way to
confine it: the sequence is $\{-1/\gamma, x,-1,f,\infty,\infty,\dots\}$ where $f$ is some finite value. We
conclude that (2.9) is not integrable.

Next we examine mapping (2.10):
$$\overline x = -1 - x(\gamma \underline x + \epsilon)\eqno(2.24)$$
If $x$=0 we find $\overline x = - 1$ and all the subsequent values are independent of the initial data:
there is no way for the mapping (2.24) to recover the lost degree of freedom and thus we conclude that
this mapping is  not integrable. Next we turn to (2.11):
$$\overline x = - 1 - {\epsilon x \over \eta \underline x + 1}\eqno(2.25)$$
and again start with $x=0$. We obtain $\overline x = - 1$, $\overline{\overline x} = - 1 +
\overline\epsilon$ and then $\overline{\overline{\overline x}}$  may become indeterminate, in the form
$0/0$, provided $\epsilon = \eta = 1$. A careful analysis of this case shows that the singularity is
periodic. Translating $x$ we find a mapping of the form:
$$\overline x \underline x + x - 1 = 0\eqno(2.26)$$
which must be integrable. It turns out that (2.26) is indeed integrable. Its integral can be readily
obtained:
$$K = {x\over{\underline x}} + {\underline x\over x} + {1\over\underline x x} - x - \underline
x-2\left({1\over x}+{1\over\underline x }\right)\eqno(2.27)$$
and (2.27) can be integrated in terms of elliptic functions. Mapping (2.26) is a member of the QRT family
of integrable mappings [9] and can be easily shown to be the only member of this family included in the
parametrization (2.5). Mapping (2.12):
$$\overline x = - 1 - {\gamma x \underline x \over \eta \underline x + 1 }\eqno(2.28)$$
has two kinds of singularities induced either by $x = 0$ or $\eta \underline x + 1 = 0$. The latter leads
to an unconfined sequence $\overline x = \infty$, $\overline {\overline x} = \infty$, etc and thus we
conclude that (2.28) is not integrable. We turn now to mapping (2.13):
$$\overline x = - x {\gamma x \underline x + \epsilon \over \delta \underline x + \zeta}\eqno(2.29)$$
A singularity starting with $\delta \underline x + \zeta = 0$ leads to an unconfined sequence $\overline
x = \infty$, $\overline{\overline x} = \infty$,$\ldots$ unless $\gamma = 0$. However even if $\gamma = 0$
we obtain a sequence $\infty$, $\infty$, finite, $0$, $0$, $\ldots$ again without recovering the
lost degree of freedom. Thus this mapping too is not integrable. Next we consider mapping (2.14). If
$\delta \neq 0$ we can put $\delta = 1$ by division (if $\delta = 0$ the mapping becomes a two point
one):
$$\overline x = - (\gamma x + \eta) - {\epsilon x\over \underline x}\eqno(2.30)$$
The singularity of (2.30) occurs when $x = 0$ and we obtain succesively $\overline x = - 1$ and
$\overline{\overline x} = \infty$. The condition for the subsequent $x$'s to be finite is just
$\overline \epsilon =\overline\gamma\underline\eta$ or $\epsilon=0$. The latter is trivial because the
mapping becomes a two-point one and we consider only the first condition. We take $\eta\ne 0$, in which
case a scaling can bring its value to $\eta=1$, resulting to $\epsilon=\gamma$.
Thus, based on singularity confinement the mapping:
$$\overline x( \gamma x + \underline x + 1) + \gamma x = 0 \eqno(2.31)$$
should be integable. This is indeed the case. Its integration can be obtained in a straightforward way.
We find as the integral of (2.31) the quantity:
$${(\overline x + 1)(x + 1)\over { a\overline x}} = K\eqno(2.32)$$
where $a$ is related to $\gamma$ through $\gamma = - {\underline a/a}$. Thus
(2.31) is the discrete derivative of a homographic mapping. Once (2.32) is obtained the complete integration of
(2.31) is elementary.

Finally, to make the long story short, we can just give without detailed proof the result that none of
the mappings (2.15), (2.17) and (2.18) satisfies the singularity confinement criterion. Thus we expect all
of them to be nonintegrable.

Before closing this section let us present the  continuous limits of the two linearizable equations we
have identified above. For (2.21) we put $x = - 1 + \nu w$, $\gamma = 3+ \nu^2 p$, $\eta = \gamma +\nu^3 q$
and we obtain at the limit $\nu\to 0$ the equation:
$$w''= 3ww'-w^3+ p w + q \eqno(2.33)$$
This is equation \#6 in the Painlev\'e/Gambier classification [10] (in noncanonical form) and precisely
the one that can be obtained from a $N=2$ projective Riccati system. For equation (2.31) we put $x = 1 +
\nu w $, $\gamma= -1-\nu^3q$ and obtain, at the limit $\nu\to 0$: 

$$w''= ww' + 2q \eqno(2.34)$$
This is equation \#5 of the  Painlev\'e/Gambier list [10] and the one resulting from the derivative of a
Riccati equation. Its canonical form can be obtained by a translation and scaling of $w$.
\vfill\eject
\noindent {\scap 3. Linearizable mappings from the Gambier systems}
\medskip\noindent
The Gambier equation is obtained as a system of two Riccati equations in cascade. (Cascade in this
context means that the first equation contains only one variable, while the second one contains
both. To solve the system one is thus led to integrate the two equations in that order, substituting the
solution of the first one into the second one). In a discrete context the Gambier system was studied in
detail  (first in [6] and in full detail in [7]). The approach is the same as in the continuous
case. One introduces two discrete Riccati equations in cascade:
$$y={b\underline y+ c\over {a\underline y+ d}}\eqno(3.1)$$
$$\overline x ={\epsilon x y + \zeta x +\eta y + \theta\over {\alpha x y + \beta x + \gamma y
+\delta}}\eqno(3.2)$$
with the obvious assumption that $ac - db\neq0$. Eliminating $y$ between (3.1) and (3.2) one obtains  a
single 3-point mapping for $x$. In this section we shall investigate the cases of the Gambier mapping that
contain the linearizable mapping. To this end we shall first reduce the Gambier equation to the form
(2.5). In practice this means that the terms quadratic in $x$ must be absent. The conditions for the
mapping to be linear in all $x$, $\overline x $,
$\underline x$ read:
$$\underline \beta (a \zeta +b\epsilon)=\underline \alpha(c\epsilon+d\zeta)$$
$$\underline \delta (a \zeta +b\epsilon)=\underline \gamma(c\epsilon+d\zeta)$$
$$\underline \alpha (c \alpha+d\beta)=\underline \beta(a\beta+b\alpha)\eqno(3.3)$$
$$\underline \gamma(c \alpha +d\beta)=\underline \delta(a\beta+b\alpha)$$
The solution of system (3.3) is long and tedious. We shall not enter into all the details but work out only
the generic case and for the remaining cases give just the results. 
The generic case is based on the assumption
that none of the terms appearing in (3.3) vanishes. By dividing the first two (or the last two) equations
we obtain $\alpha \delta=\beta \gamma$. We then form the product of the second and third equation, use the
previously derived  relation and expanding we find
$\alpha\zeta=\beta\epsilon$. It suffices then to substitute into any of the four equations in order to get
a final relation. The nongeneric cases are obtained by assuming that specific terms do vanish. Five cases
can be distinguished, finally:
\medskip
\item{1)}
$  \zeta={\displaystyle \epsilon \beta\over \displaystyle\alpha},\
       \beta={\displaystyle \alpha \delta\over \displaystyle\gamma},\
       c={\displaystyle a\delta
\underline\delta+b\underline\delta\gamma-d\delta\underline\gamma\over \displaystyle\gamma\underline\gamma}$

\item{2)}
$  \alpha=\beta =0,\   \underline \delta={\displaystyle \underline \gamma
( c\epsilon+ d\zeta)\over \displaystyle a
\zeta + b\epsilon}$

\item{3)}
$\alpha=\beta =\gamma=0,\  a \zeta +b\epsilon=0$

\item{4)}
$\alpha =\gamma=\epsilon=a=0$

\item{5)}
$\delta =\gamma=0,\  \zeta={\displaystyle\epsilon \beta\over \displaystyle\alpha}$,\ 
    $c={\displaystyle\underline\beta
(a\beta+b\alpha)-d\beta\underline\alpha\over\displaystyle\alpha \underline\alpha}$
\medskip 
Once these basic conditions are obtained we still have full homographic freedom which will allow us to reduce
further (3.1) and (3.2). Let us work out in detail the case corresponding to constraint 1)
above. First we can translate $y$ so as to bring $\delta$ to zero. The direct consequence
of $\delta=0$ is $\zeta=\beta=c=0$. We can further translate $x$ so as to get $\gamma=0$
and by division put $b=1$, $\alpha=1$. Finally by scaling $x$ we can take $\epsilon=1$, unless $\epsilon=0$
and by scaling  $y$ we can put $\theta=1$. In the case $\epsilon=0$ we can use the scaling of $x$ to put
$\eta=1$ (unless, of course, $\eta=0$) and then put $\theta=1$. As a result of these transformations case
1) can be reduced to three cases:
$$y={\underline y\over a \underline y + d}\eqno(3.4)$$
combined with one of the three equation for $x$ below:
$$\overline x={x y+\eta y + 1\over x y }\eqno(3.5)$$
or
$$\overline x={y + 1\over x y }\eqno(3.6)$$
or
$$\overline x={1\over x y }\eqno(3.7)$$
Concerning the remaining cases we just give the final results. Case 2) leads to the
following mapping:
$$y={b\underline y-1\over a \underline y + 1}\eqno(3.8)$$
$$\overline x={x y+x+ \theta\over  y }\eqno(3.9)$$
which is the main type and also to three simpler cases
$$y={b\underline y\over a \underline y + 1}\eqno(3.10)$$
$$\overline x={x y+1\over  y }\eqno(3.11)$$
and also:
$$y={b\underline y+c\over  \underline y }\eqno(3.12)$$
coupled to either:

$$\overline x={x+ y\over  y }\eqno(3.13)$$
or
$$\overline x={x \over  y }\eqno(3.14)$$
Case 3) leads to cases identical to some of the ones identified in case 2) plus one trivial
case corresponding to a purely linear mapping $y=b\underline y+c$, $\overline x=x+y$.

Case 4) yields the mapping:

$$y=b\underline y+c\eqno(3.15)$$
$$\overline x={\zeta x+ y\over x }\eqno(3.16)$$
Finally case 5) is identical to case 1). We can readily remark that the nongeneric mapping (3.4) or (3.10) is 
in fact linear for $1/y$. Thus the system (3.10), (3.11) can be readily discarded as
been linear. Moreover (3.15), (3.16) is a subcase of (3.4), (3.5) (once $y$ is inverted).
We can thus summarize the cases to be studied in detail in the following list (where we
have inverted $y$ in (3.4)). We obtain just three cases, where
$\sigma=0$ or 1.

{I)}
$$y=a+d\underline y,\qquad x \overline x=\sigma x+y \eqno(3.17)$$
\noindent (obtained from case 1) up to a translation of $y$).

{II)}
$$y=b+{c\over\underline y},\qquad \overline x={x\over y}+\sigma \eqno(3.18)$$
\noindent (obtained from the simplified case 2)) and

{III)}
$$y={b\underline y-1\over a \underline y + 1},\qquad\overline x={x y+x+ \sigma\theta\over  y }\eqno(3.19)$$
(obtained from the main type of case 2)).
The cases with $\sigma=0$ are trivially integrable: once $y$ is given the integration for
$x$ is obtained through a multiplicative equation of the form $\overline x x=F(n)$ or
${\overline x/ x} = F(n)$, which can be linearized by just
taking the logarithm of both sides.

In order to investigate the integrability of the $\sigma=1$ cases we shall apply the singularity
confinement criterion. In the case of the mapping (3.17) a singularity appears when $y=0$ in which case
$\overline x$ does not depend on the value of $x$. No way exists for the mapping to retrieve this lost
degree of freedom through some indeterminate form. Thus (3.17) for $\sigma=1$ is not integrable.
We consider now the mapping (3.18) for $\sigma=1$:
$$y=b+{c\over\underline y}\eqno(3.20a)$$
$$\overline x={x\over y}+1 \eqno(3.20b)$$
The sequence of singular values of (3.20a) is $y=0$, $\overline y=\infty$ and finite values for
$\overline{\overline y}$ and beyond. The fact that $y=0$ induces a divergence on
$\overline x$, and at leading order we have $\overline x\propto {x/y}$. At the next
step we must take into account that $\overline y\propto {\overline c/y}$ and thus
the computation of $\overline{\overline x}$ leads to 
$\overline{\overline x} = 1+{ x/ \overline c}$ that is a finite value
depending on the initial condition $x$. Thus the singularity is confined and the mapping
(3.20) is expected to be integrable. 
Finally, we examine mapping (3.19) for $\sigma=1$. The only singularity sequence is 
$\underline y=1/b$, $y=0$, $\overline y=-1$ etc. Starting with a regular $\underline x$ we find a finite
value for $x$, while $\overline x$ diverges. However a close examination of the singularity shows that 
$\overline {\overline x}$ is finite and the singularity is confined without any constraint. (Moreover, the
special value $a=1$, which leads to a more complicated singularity pattern for $y$, also has confined
singularities for $x$). Thus the mapping (3.19) is expected to be integrable in general. Before proceeding
further let us transform (3.19) to a form close to that of (3.20). It is possible to show that by
translating $x$, $x\to x-\theta$, and a homographic transformation on $y$, $y\to y/(1-y)$, we can bring
(3.19) to the form:
$$y=A+{B\over\underline y}\eqno(3.21a)$$
$$\overline x={x\over y}+\rho\eqno(3.21b)$$
where $A=(b+1)/(a+b)$, $B=-1/(a+b)$ and $\rho=\overline\theta-\theta$.

We turn now to the integration of mappings (3.20) and (3.21).
Instead of restricting ourselves to the particular forms of (3.20) and (3.21) we shall generalize
our setting a little since this will lead to interesting results. Let us start with the
more general Gambier system:
$$y=a+{b\over\underline y}\eqno(3.22a)$$
$$\overline x={x\over y}+{p y +q\over ry+s}\eqno(3.22b)$$
The application of singularity confinement follows exactly the same steps as for (3.20-21)
and shows that (3.22) satisfies this integrabilty requirement. Thus (3.22) is expected
to be integrable. Next we can ask for which values of $p$, $q$, $r$, $s$ is the 3-point
mapping for $x$ is of the linearizable form (2.5). It turns out that this is possible
only if the equation for $x$ of the form:
$$\overline x = {x \over y}+{\theta - \phi y\over y}\eqno (3.23)$$
If $\phi=\overline \theta$ then it is possible to bring them both to zero (by tranlation
of $x$), and the mapping becomes trivial. Otherwise it is possible, through a
translation, to bring either $\phi$ or $\theta$ to zero and scale the remaining one to
unity. Thus the only interesting forms of (3.23) are:
$$\overline x={x+1\over y} \eqno(3.24)$$
or (with $\xi=x+1$) 
$$\overline \xi={\xi\over y}+1 \eqno(3.25)$$
 One introduces the transformation $x=- (1+\omega X)^{-1}$  (or equivalently
 $\xi=\omega X(1+\omega X)^{-1}$) where
$\overline\omega \omega \underline \omega=-b$ and reduce the 3-point mapping resulting
from the elimination of $y$ between (3.22a) and (3.24) to: 
$$\overline X X  \underline X+ \underline X X{a+1\over \overline \omega}+ \underline
X \ \underline \omega(1-{a\over b}) +1=0\eqno(3.26)$$
 We remark that (3.26) is the generic linearizable mapping presented in its fully reduced
form in (2.21). Thus the Gambier mapping leads again to the linearizable one obtained
from the projective discrete Riccati system is perfect analogy to the continuous case.
\bigskip
\noindent {\scap 4. Conclusion}
\medskip
\noindent 
In the previous sections we have investigated 3-point mappings that are integrable
through linearization. Our analysis was guided by the analogy with the continuous
situation and results of ours on $N=3$ projective systems and the Gambier equation. We
have presented a exhaustive analysis of the linearizable mapping and identified all its
integrable forms. We thus have shown that the discrete equation:
$$\overline X X  \underline X+\alpha \underline X X+\beta \underline X  +1=0\eqno(4.1)$$
can be reduced to a linear $3 \times 3$ system. The same equation, with the transformation
$X=c {(1+1/ x)}$ can be reduced to the Gambier system:
$$y =a+{b\over \underline y}\eqno(4.2)$$
$$\overline x = {x+1\over y}$$
(where $a$, $b$, $c$ are related to the $\alpha$, $\beta$), which provides a different method for its
solution.
Nongeneric cases of the general linearizable mapping were also identified and in particular the equation:
$$\overline x (\gamma x + \underline x +1) + \gamma x =0\eqno(4.3)$$
that is the discrete derivative of a homographic mapping and constitute thus the discretisation of
equation \#5 on the  Painlev\'e/Gambier  classification. The extension of the present
results to higher order (third and beyond) systems could be most interesting. We intend
to return to this question in some future work.

 \bigskip
\noindent {\scap Acknowledgements}.
\smallskip
\noindent 
The financial help of
the CEFIPRA, through the contract 1201-1, is gratefully acknowledged. S. Lafortune acknowledges two
scholarships: one from NSERC (National Science and Engineering Research Council of Canada) for his Ph.D.
and one from ``Programme de Soutien de Cotutelle de Th\`ese de doctorat du Gouvernement du Qu\'ebec'' for
his stay in Paris.
\bigskip
{\scap References}
\smallskip 
\item{[1]} M.D. Kruskal, A. Ramani and B. Grammaticos, NATO ASI Series C 310, Kluwer
1989, 321.
\item{[2]} R.L. Anderson, J. Harnad and P. Winternitz, Physica D4 (1982) 164-182.
\item{[3]} B. Grammaticos, A. Ramani and P. Winternitz, {\sl Dicretizing families of
linearizable equations}, preprint (1997).
\item{[4]} A. Ramani, B. Grammaticos and G. Karra, Physica A 181 (1992) 115.
\item{[5]} B. Gambier, Acta Math. 33 (1910) 1.
\item{[6]} B. Grammaticos and A. Ramani, Physica A 223 (1995) 125.
\item{[7]} B. Grammaticos, A. Ramani and S. Lafortune, {\sl The Gambier mapping,
revisited}, in preparation.
\item{[8]} B. Grammaticos, A. Ramani and V. Papageorgiou, Phys. Rev. Lett. 67 (1991)
1825.
\item{[9]} G.R.W. Quispel, J.A.G. Roberts and C.J. Thompson, Physica D34 (1989) 183.
\item{[10]} E.L. Ince, {\sl Ordinary differential equations}, Dover, New York, 1956.

 \end